\documentclass[twocolumn,showpacs,preprintnumbers,amsmath,amssymb]{revtex4}
\usepackage{graphicx}
\usepackage{dcolumn}
\usepackage{bm}

\def\>{\rangle}
\def\<{\langle}

\def\ave#1{\left\< #1\right\>}

\def\dd{{\rm d}}

\begin{document}

\title{Universal decay of classical Loschmidt echo of
neutrally stable but mixing dynamics}

\author{Giulio Casati$^{1,2,3}$, Toma\v z Prosen$^{4,3}$, Jinghua Lan$^3$ and 
Baowen Li$^3$}
\affiliation{
$^1$Center for Nonlinear and Complex Systems, 
Universita' degli studi dell'Insubria, Como\\
$^2$Istituto Nazionale di Fisica della Materia, Unita' di Como, and \\
Istituto Nazionale di Fisica Nucleare, sezione di Milano, 
Milano, Italy\\
$^3$Department of Physics, National University of Singapore,  Singapore 117542, Republic of Singapore
\\
$^4$ Physics Department, Faculty of Mathematics and Physics, University of 
Ljubljana, Ljubljana, Slovenia }

\date{\today}

\begin{abstract}
We provide analytical and numerical evidence that classical mixing systems 
which lack exponential 
sensitivity on initial conditions, exhibit universal decay of Loschmidt echo
which turns out to be a function of a single scaled time variable $\delta^{2/5}t$,  
where $\delta$ is the strength of perturbation. 
The role of dynamical instability and entropy production is discussed.

\end{abstract}


\maketitle

Fidelity, or Loschmidt echo, is defined as the overlap
of two time evolving states which, starting  from the same initial condition, evolve under two slightly different Hamiltonians. It is therefore an important quantity which measures the stability of the motion under system's perturbations.
The recent interest in the behaviour of fidelity \cite{Peres,Usaj,Jalabert,Prosen,Beenakker,Tomsovic,PZ,BC1,Eckhardt,Veble,VP} has been largely
motivated by a possible use in quantifying stability of quantum computation \cite{Qcomp}.

It has been shown \cite{Veble} that for classical chaotic, exponentially unstable systems, the decay rate of fidelity is {\it perturbation independent} and, asymptotically, fidelity decays as correlation functions.
On the other hand, for quantum systems, fidelity decay obeys different regimes {\it depending} on perturbation strength.
In this relation, particularly intriguing is the recently discovered case of mixing dynamics with vanishing Lyapounov exponent \cite{caspro99,caspro00} a prominent example of which are billiards inside polygons \cite{note}.
In several respects the classical dynamics of such systems
is reminiscent of quantum dynamics of generic chaotic systems which, apart from an initial time,  exponentially short in ${\hbar}$, are linearly stable. As a consequence statistical relaxation in quantum mechanics takes place in absence of exponential instability. Certainly, the dramatic difference in the dynamical stability properties of different systems must be reflected in a different qualitative behaviour of physical quantities such as fidelity which is the object of the present paper.

In the following, under the assumptions of linear separation
of trajectories and dynamical mixing, which may be 
produced by some discontinuity in the flow,
we derive a universal scaling law of classical fidelity decay.
We conjecture that this surprising fidelity decay may be associated to a peculiar
power-logarithmic entropy production in such systems.
We consider here a specific example, i.e. the triangle map $z_{n+1}=T(z_n)$ \cite{caspro00} on 
a torus 
$z=(x,y)\in [-1,1)\times [-1,1)$
\begin{eqnarray}
y_{n+1} &=& y_n + \alpha {\rm\, sgn} x_n + \beta \pmod{2}, \nonumber \\
x_{n+1} &=& x_n + y_{n+1} \pmod{2}, \label{eq:map1}
\end{eqnarray}
where ${\rm sgn}x=\pm 1$ is the sign of $x$ and
$\alpha,\beta$ are two parameters. Previous investigations have shown
that \cite{caspro00} (see also \cite{mirko} for some rigorous results on 
(\ref{eq:map1})): for rational values of $\alpha,\beta$ the system is pseudo-integrable,
as the dynamics is confined on invariant curves. If $\alpha=0$ and $\beta$ is irrational, the dynamics
is (uniquely) ergodic, but not mixing, while for
incommensurate irrational values of $\alpha,\beta$ the dynamics is ergodic and mixing with dynamical
correlation functions decaying as $t^{-3/2}$. It can be argued that the triangle map possesses the
essential features of bounce maps of polygonal billiards and 1d hardpoint gases \cite{caspro99,caspro00}, namely parabolic stability in 
combination with decaying dynamical correlations, and as such 
represents a paradigmatic model for a larger class of systems.

The classical fidelity
$F_\delta(n)$ can be written as an overlap of two 
phase space densities propagated by the original map $T$ and the perturbed map 
$T_\delta = T \circ g_\delta$ where $g_\delta(z) = z + \delta a(z)$ is some near-identity area-preserving map 
parametrized by a vector field $a(z)$:
\begin{equation}
F_\delta(n) = \frac{\int \dd^2 z \rho(T^{(-n)}(z)) \rho(T^{(-n)}_\delta(z))}{\int \dd^2 z \rho^2(z)},
\label{eq:fidelity}
\end{equation}
We can make our discussion even more general by
taking the perturbation {\em explicitly time-dependent}. Let the perturbed map $T_{\delta,n}$ explicitly depend
on iteration time, namely we consider the following class of perturbed triangle maps,
$\bar{z}_{n+1} = T_{\delta,n}(\bar{z}_n)$
\begin{eqnarray}
\bar{y}_{n+1} &=& \bar{y}_n + \alpha {\rm\, sgn} \bar{x}_n + \beta + \delta f_n(\bar{x}_n) \pmod{2}, \nonumber \\
\bar{x}_{n+1} &=& \bar{x}_n + \bar{y}_{n+1} \pmod{2}. \label{eq:mapp}
\end{eqnarray}

\begin{figure}
\includegraphics[width=\columnwidth]{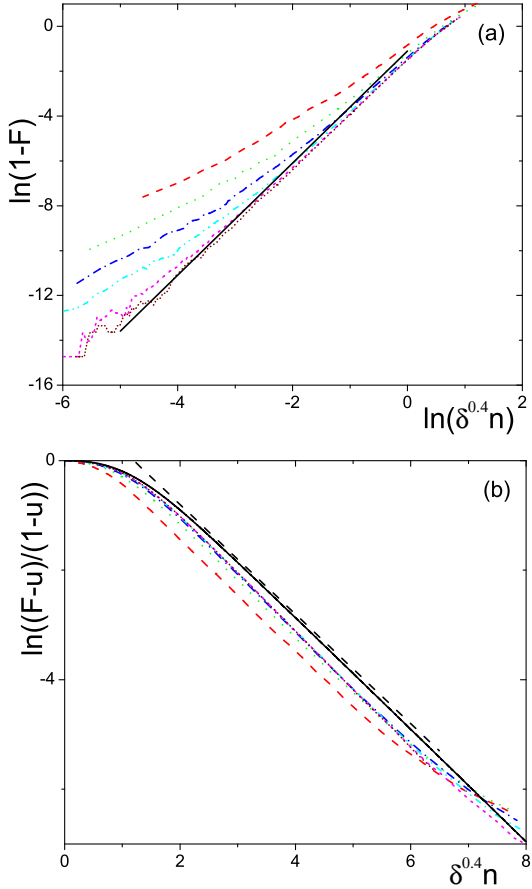}
\vspace{-1cm}
\caption{\label{fig:fn}Fidelity decay. To compute fidelity we divide the phase
space in $100\times 100$ cells. We then take $10^5$ points in one cell and 
evolve these points with the map (\ref{eq:map1}) up to time $n$. Then we 
compute the reverse evolution with the map (\ref{eq:mapp}) and compute the
fraction of points which fall again in the initial cell, after time $2n$. The
result is then averaged over initial distributions in 49 different randomly chosen cells. (a) $\log(1-F_\delta(n))$ versus $\log(\delta^{2/5} n)$ magnifying behaviour for short times. The broken curves refer to 
$\delta=10^{-5},\ldots,10^{-10}$ (top-down). The full line is the theoretical expression (\ref{eq:shortt}). (b) $\log[(F_\delta(n)-u)/(1-u)]$ versus
$\delta^{2/5} n$ magnifying asymptotic (long-time) behavior. Here $u=10^{-4}$ is the relative area of the initial set ${\cal A}$ giving the 
asymptotic value of fidelity. 
The meaning of broken curves is the same as in (a). The full curve gives
the numerical solution of the random Gaussian model (\ref{eq:rm}), while the
dashed line has slope $-1$ to indicate asymptotic exponential decay.}
\end{figure}
We will assume that the {\em force function} $f_n$ has {\em vanishing time-average} for almost 
any initial condition.
Let as further assume that the initial density $\rho(z)$ is a characteristic function over some set ${\cal A}$ of 
typical diameter $\omega$ with 
$\delta \ll \omega \ll 1$. Then a pair of orbits $z_n$ and $\bar{z}_n$ 
starting from the same point $\bar{z}_0=z_0$ in ${\cal A}$, contribute to
(\ref{eq:fidelity})
until they hit the opposite sides of the discontinuity, at $x=0,1 \pmod{2}$.
The fidelity at time $n$ is then simply the probability that the pair of orbits does not hit the cut up to $n-$th iterate. Assuming ergodicity of the map \cite{caspro00} we write
\begin{equation}
F_\delta(n) = \ave{\prod_{n'=1}^n (1 - |\Delta x_{n'}|)}
\label{eq:prodfid}
\end{equation}
where $\Delta x_n = \bar{x}_n - x_n$, and $\ave{A_n} = \int \dd z \rho(z) A(T^{(n)}(z))/\int \dd z \rho(z)$.
In order to derive the fidelity decay for the triangle map we have to 
compute the
average growth rate of the orbits distance perpendicular to the cut.
This is achieved by writing out an explicit linearized map for the orbits 
displacement $\Delta z_n = \bar{z}_n - z_n$
\begin{equation}
\Delta z_{n+1} = 
\left(
\begin{array}{cc}
1 & 1 \\
0 & 1 \\
\end{array}\right)\Delta z_n + 
\left(
\begin{array}{c}
1 \\ 
1 \\
\end{array}\right)\delta f_n(x_n).
\end{equation}
This system of linear difference equations can be solved explicitly, say for $\Delta x_n$: 
\begin{equation}
\Delta x_n = \delta \sum_{n'=0}^{n-1} (n-n') f_{n'}(x_{n'})
\label{eq:expl}
\end{equation} 
with the initial condition $\Delta z_0 = 0$. Assuming that $f_n$ are pseudo-random variables with quickly
decaying correlation function $C(n) = \lim_{m\to\infty}\ave{f_{m}(x_m) f_{n+m}(x_{n+m})}$, we can employ a version
of the {\em central limit theorem} to show that $\Delta x_n$ should have {\em Gaussian distribution} for sufficiently
large $n$. To this end, let us first notice that the second moment 
\begin{equation}
\ave{(\Delta x_n)^2} = \delta^2 \sum_{n'=0}^{n-1}\sum_{n''=0}^{n-1} (n-n')(n-n'')\ave{f_{n'}f_{n''}}
\end{equation}
can be related, as $n\to\infty$, to the integrated correlation function. 
Since, for large $n$: $\ave{f_{n'}f_{n''}} = C(n'-n'')$, we obtain by means of a straightforward calculation
\begin{equation}
\ave{(\Delta x_n)^2} \longrightarrow \frac{1}{3}\delta^2 n^3 \sigma,\quad
\sigma := \sum_{m=-\infty}^\infty C(m).
\label{eq:varian}
\end{equation}
Now, as long as fidelity remains close to $1$, we can expand (\ref{eq:prodfid})
to first order 
$F_\delta(n) = 1 - \sum_{n'=1}^n \ave{|\Delta x_{n'}|},$
where the average 
$\ave{|\Delta x_n|} = \sqrt{2\sigma/(3\pi)}|\delta| |n|^{3/2}$
can be computed using a Gaussian distribution of 
$\Delta x_n$ with variance $\ave{(\Delta x_n)^2}$ given in (\ref{eq:varian}). 
This yields
\begin{equation}
F_\delta(n) = 1 - \sqrt{\frac{8\sigma}{75\pi}} |\delta| |n|^{5/2}. 
\label{eq:shortt}
\end{equation}
This expression is valid until $F_\delta(n)$ remains close to 1,
that is up to time $|n| < n^* = \sigma^{-1/5} |\delta|^{-2/5}.$


%
In fig.~1a we show the behavior of $1-F_\delta(n)$ for short times $n < n^*$ and compare with the
theoretical formula (\ref{eq:shortt}) with $\sigma=3.29 \pm 0.01$ as 
computed from numerical simulation of correlation function 
$C(n)$. As for perturbation we choose a simple shift in the parameter $\alpha$, so the force reads $f(x) = {\rm sgn}\, x$ and is, in this case, {\em not explicitly} time dependent. Yet it is pseudorandom and one can see that, 
as $\delta$ decreases, the numerical curves approach the theoretical 
expression (\ref{eq:shortt}). 

Notice that according to eq.~(\ref{eq:varian}), the average distance between
two orbits increases as $\propto n^{3/2}$. On the other hand, the distance between two
initially close orbits of the {\em same} map increases only linearly with
time.
This is nicely confirmed by the numerical simulations of fig.~2. 

\begin{figure}
\includegraphics[width=7.5cm]{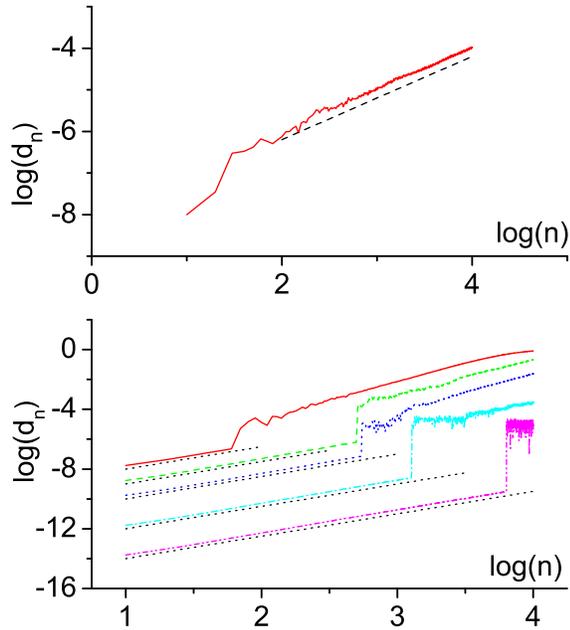}
\caption{\label{fig:distance}(a) Average distance 
$d_n = |\Delta z_n|$
 versus time $n$ for two nearby initial orbits of the unperturbed map (\ref{eq:map1}). 
The initial distance is $d_0=10^{-9}$, the average is taken over $2.5\times 10^8$ different initial conditions. 
The dotted line has slope 1. (b) Average distance versus time for two orbits starting from the same initial condition
and evolving under the unperturbed and perturbed maps, (\ref{eq:map1}) and (\ref{eq:mapp}), respectively.
The values of perturbation $\delta$, for the curves (from top down), are: 
$10^{-9}, 10^{-10}, 10^{-11}, 10^{-13}$, $10^{-15}$.
The data are averaged over $10^5$ different initial conditions. Dotted lines have slope $1.5$.
}
\end{figure}

\begin{figure}
\includegraphics[width=0.95\columnwidth]{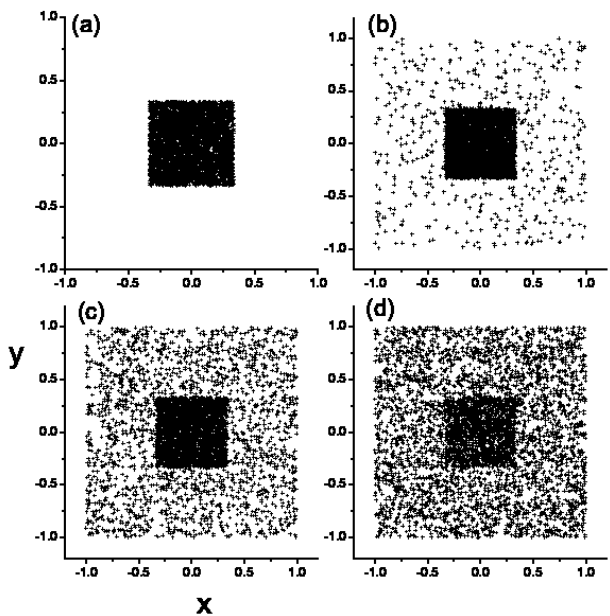}
\vspace{-3mm}
\caption{\label{fig:spreading}
Spreading of phase space points of the echo-dynamics.
We consider 5000 initial points in the central cell of a
$3\times 3$ grid (a). We then evolve these points up to time $n$, and then reverse the motion
with the perturbed dynamics, with $\delta=10^{-6}$, 
up to the echo time $2 n$. The density of points, for $n=200,400,600$ 
is shown in (b),(c),(d), respectively.
}
\end{figure}

For larger times, $n > n^*$, higher order terms in 
the expansion of (\ref{eq:prodfid}) contribute, so temporal 
correlations among $\Delta x_n$ become important. We are here unable to derive 
exact theoretical predictions for the 
fidelity decay in this regime. However, numerical
results in fig.~1b show that, for large times,
fidelity decays exponentially 
$F_\delta(n) = \exp(-\gamma |n|)$ with exponent $\gamma = C |\delta|^{2/5}$.
We also checked that the transition time between the two regimes of decay, 
scales as $\delta^{-2/5}$. In conclusion, extensive and accurate numerical 
results provide clear evidence that fidelity depends on the single scaling
variable $\tau = \delta^{2/5} n$. 

In the following, we show that this scaling behavior can be derived analytically for sufficiently small $\delta$.
The only assumption is correlation decay with a {\em finite} characteristic 
time-scale $n_{\rm mix}$, i.e. $\ave{f_n f_{n'}}$ practically vanish for $|n-n'|> n_{\rm mix}$.
Let us divide the time-span $n$ into $\nu:=n/m$ blocks of $m$ steps each, 
such that $n_{\rm mix} \ll m \ll n$,, and make a scaling argument. 
The local variation of $\Delta x_n$, namely 
$\Delta x_{n+1} - \Delta x_{n} = \delta\sum_{n'=0}^{n} f_{n'} \sim \delta \sqrt{n}$
is much smaller than the mean value $\ave{|\Delta x_n|} \sim \delta n^{3/2}$.
Thus we approximate the product (\ref{eq:prodfid}) within each block labelled 
by $\iota=1,\ldots \nu $ as
$(1 - |\Delta x_{(\iota-1)m}|)^m \approx 1 - m |\Delta x_{(\iota-1)m}|$. Therefore
\begin{equation}
F_\delta(n) \approx \prod_{\iota=1}^{\nu} (1 - m |\Delta x_{(\iota-1)m}|).
\label{eq:renfid}
\end{equation}
Next we define the normalized block-averaged forces 
\begin{equation}
\xi_\iota = \frac{1}{\sqrt{\sigma m}}\sum_{k=0}^{m-1} f_{(\iota-1)m + k}
\end{equation}
which are normalized, and uncorrelated, $\ave{\xi_\iota \xi_\mu} = \delta_{\iota \mu}$ 
since $m \gg n_{\rm mix}$. Using Eq.(\ref{eq:expl}) we can write
$\Delta x_{(\iota-1)m} \approx \delta \sum_{\mu=1}^\iota (\iota-\mu)m 
\sum_{k=0}^{m-1} f_{(\mu-1)m + k}
= \delta m^{3/2} \sigma^{1/2} \sum_{\mu=1}^\iota (\iota-\mu) \xi_\mu$. 
If, in additon to the rescaled time $\nu=n/m$, we define a rescaled perturbation 
$\epsilon=\delta \sigma^{1/2} m^{5/2}$ then we can write Eq. (\ref{eq:renfid}) as
\begin{equation}
\Phi_\epsilon(\nu) = \ave{ \prod_{\iota=1}^\nu \left(1 - \left|\epsilon 
\sum_{\mu=0}^{\iota-1} (\iota-\mu)
\xi_{\mu}\right|\right)}_{\xi}.
\label{eq:rm}
\end{equation}
The derived relation
$
F_{\delta}(n) = \Phi_{\delta \sigma^{1/2} m^{5/2}}(n/m)
$
does not depend on $m$ (for large enough $m$), 
and therefore fidelity should be a function of the scaling 
variable $\tau = |\delta|^{2/5} n$ only. 

Notice that due to the central limit theorem, 
since $m\gg n_{\rm mix}$, $\xi_\mu$ can be simply treated as uncorrelated, 
normalized, Gaussian stochastic variables. We have actually computed the {\em universal function} 
$\phi(\epsilon^{2/5} \nu) = \Phi_\epsilon(\nu)$ by means of Monte-carlo integration, and checked that
it is practically insensitive to $\epsilon$, for $\epsilon < 10^{-4}$.
As it is seen in fig.~1b, the numerical data for the triangle map
agree with the theoretical expression (\ref{eq:rm}),
namely $\phi(\delta^{2/5}\sigma^{1/5} n)$ which is plotted as a full curve.

The two regimes of fidelity decay described above are illustrated in fig.~3 
by the image at the echo time of an initial uniform phase space distribution over some set 
${\cal A}$. Notice that the linear-response 
regime (\ref{eq:shortt}) is valid until the shape of the intial set is approximately restored
at the echo time. For larger times, the fidelity decay becomes exponential.

\begin{figure}
\includegraphics[width=\columnwidth]{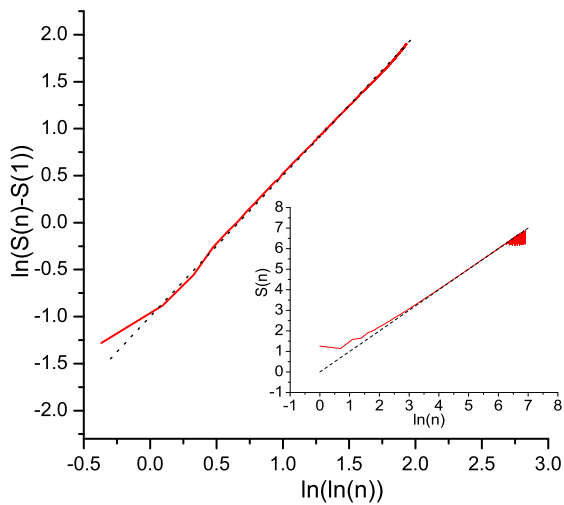}
\vspace{-1cm}
\caption{
The time evolution of the coarse-grained entropy for the triangle map,
computed by taking $5\times 10^7$ points initially 
distributed randomly in one cell of a
$N\times N$ phase-space grid with $N=700$. 
We plot $\ln(S(n)-S(1))$ versus $\ln\ln n$.
The straight line has slope $3/2$. Obviously, the entropy 
will eventually saturate at $S(\infty)=\ln N^2$.
In the inset we show the entropy computed for the 
uniquely-ergodic and nonmixing dynamics with
$\alpha=0$. The dotted line has slope $1$.
}
\end{figure}

Finally we would like to stress that this behavior of the triangle map differs from the
typical behavior which has been found for chaotic or for integrable systems.
In particular, contrary to the case
of exponentially unstable systems, in this case the rate of fidelity decay
depends on the perturbation strength. This feature is shared by quantum
systems in which exponential instability is absent as well.
One may wonder if this behavior is reflected also in some other, perhaps even more fundamental
dynamical property of the map. In order to explore this question, we have computed 
the entropy production for the map (\ref{eq:map1}). 
As the extensive computation of Kolmogorov-Sinai 
dynamical entropies seemed too expensive for reaching any conclusive results, we have decided
to compute the dynamical evolution of the coarse grained statistical entropy 
$S_n = -\sum_j p_n^{(j)} \ln p_n^{(j)}$.
To this end we divide the phase space in $N\times N$ equal cells, and consider
an initial ensemble of points uniformly distributed over one cell.
The probability $p_n^{(j)}$ is defined as the fraction of orbits which, after $n$ time steps,
fall in the cell of label $j$.
For a chaotic system with dynamical entropy $h$, one expects $S_n = h n + {\rm const}$ \cite{latora},
whereas for ergodic-only (non-mixing) dynamics one expects $S_n \sim \ln n$, for sufficiently large $N$.
Our numerical results for the triangle map (fig.~4) show instead that
$S_n-S_1 = |\ln n|^\lambda$ 
with the exponent $\lambda = 3/2$. 
Furthermore, as shown in the inset of fig.4,
for the triangle map (\ref{eq:map1}) with $\alpha=0$ numerical results give,
quite accurately, $S(n) = \ln n$ (with no prefactor or additional constant).

In conclusion, we have discussed the parametric stability, as 
characterized by classical fidelity or Loschmidt echo, of an important class of 
dynamical systems where neutral stability is coexisting with dynamical mixing.
As a paradigmatic example of this class of systems we have considered the triangle map. 
By means of analytic calculations and numerical simulations we have derived two universal 
regimes of fidelity decay, both being characterized by a universal 
scaled time variable $|\delta|^{2/5}t$. This interesting dynamical behavior is 
supported also by a power-logarithmic behavior of the coarse-grained entropy.

We acknowledge financial support by the PRIN 2002 
``Fault tolerance, control and stability of
quantum information processing'' and PA INFM ``Weak chaos: Theory and 
applications'' (GC), by the grant P1-044 of
 Ministry of Education, Science and Sport of Slovenia (TP), by the grant 
DAAD19-02-1-0086, ARO United States (GC and TP),
and by the Faculty Research Grant of NUS and the Temasek
Young Investigator Award of DSTA, Singapore under project agreement
POD0410553 (BL).


\begin{thebibliography}{99}
\bibitem{note} Strictly speaking there is no rigorous proof that irrational polygons are mixing. We refer here to the evidence provided by numerical computations such as those reported in refs.~\cite{caspro99,caspro00}.
\end{thebibliography}
\end{document}